# Features of oxygen and its vacancies diffusion in $YBa_2Cu_3O_{7-\delta}$ thin films near to magnetic quantum lines

Viktor O. Ledenyov, Dimitri O. Ledenyov and Oleg P. Ledenyov
National Scientific Centre Kharkov Institute of Physics and Technology,
Academicheskaya 1, Kharkov 61108, Ukraine.

The temperature dependence of the real and imaginary parts of magnetic susceptibility of $YBa_2Cu_3O_{7-\delta}$ superconducting thin films were researched at the electron irradiation with the energy of $2 \cdot 10^4 \ eV$ in the magnetic field of $\sim 2.4 \cdot 10^4 \ A/m$ at the temperature $T = 77 \ K$. It is retrieved that the nonequilibrium distribution of atoms of oxygen, originating at the operation of an electron beam, reduces in the change of a sort of dependence of the complex magnetic susceptibility of $YBa_2Cu_3O_{7-\delta}$ thin films on the temperature $T$. The diffusion relaxation of a none equilibrium distribution of oxygen atoms and its vacancies, generated at the application of electron irradiation, results in an appearance of pinning centers with the reduced contents of the oxygen in the normal cores of the Abricosov magnetic vortices. The increase of pinning forces results in the increase of critical current density, having an effect on the magnetic susceptibility of $YBa_2Cu_3O_{7-\delta}$ thin films. The proposed theoretical mechanism on the oxygen and its vacancies diffusion in $YBa_2Cu_3O_{7-\delta}$ superconducting thin films near to the centers of Abricosov magnetic vortices cores at the action of the superconducting electron pair potential gradient explains the observed physical properties.

## Introduction

The oxygen is an important chemical element in the $YBa_2Cu_3O_{7-\delta}$ high-temperature superconductor crystal grating, which defines its electronic, magnetic and superconducting properties. The change of the oxygen content from the index value of $7$ to $6.5$ results in the decrease of $YBa_2Cu_3O_{7-\delta}$ critical transition temperature from $T_C \approx 93 \ K$ to $T_C = 0 \ K$, transforming it to the dielectric state with an appearance of antiferromagnetic properties. In view of strong influence by the oxygen on the electronic properties of $YBa_2Cu_3O_{7-\delta}$, the processes of the saturation of crystal grating by the oxygen due to its diffusion in a crystal (In-Diffusion) and the oxygen outgo from a crystal (Out-Diffusion) were researched in details [1, 2]. The dependence of oxygen diffusion coefficients in a wide range of temperatures was also investigated. The obtained results show that:

i. The oxygen diffusion is not featured by one general purpose coefficient with the uniform value of activation energy;

ii. The oxygen diffusion coefficients have different values at the high and low temperatures.

Thus, the measurements of $YBa_2Cu_3O_{7-\delta}$ high-temperature superconducting compounds show that the activation energy $E_A$ of oxygen diffusion in the **ab** plane in a crystal grating is close to $1 \ eV$, and in some cases, it is equal to $0.4 \ eV$ [1]. The values of parameters of oxygen diffusion in the **b** axis direction can be even smaller [3]. Let's note that, in the oxide semiconductors (*OS*) [4], the presence of the two activation energies is characteristic. The high-temperature oxygen diffusion has the big value of the activation energy $E_A$. Thus, at this situation, the equilibrium balance between the crystal grating and the oxygen is shifted toward the side of chemical composition with the small level of oxygen contents in the *OS*. Whereas, at low temperatures, the transition to the higher oxygen index with the smaller value of activation energy is observed in the *OS*. Probably, the similar physical properties may be characteristic for the high temperature superconducting (*HTS*) compounds, though this problem, as it is mentioned by different authors, has not been researched in details. The oxygen can leave the $YBa_2Cu_3O_{7-\delta}$ compounds at application of heating. The annealing of $YBa_2Cu_3O_{7-\delta}$ in the oxygen atmosphere at the moderately low temperature of $720 \ K$ is necessary for the oxygen saturation purposes.

This research is focused on the consideration of the nature of the oxygen and its vacancies diffusion in $YBa_2Cu_3O_{7-\delta}$ thin films in close proximity to the Abricosov magnetic vortices normal cores. The diffusion mobility of the all other atoms is very small in comparison with the oxygen atoms [2]. The oxygen diffusion process can be originated by the Abricosov magnetic vortical nanostructures, penetrating into the superconductor at the certain magnitude of applied magnetic field: $H_e > H_{c1}$ [5], where $H_{c1}$ is the low critical magnetic field of the superconductor. The Abricosov magnetic vortex has a normal metal core with the characteristic radius $\sim \xi$, where $\xi$ is the coherence length of a superconductor. There is a conditional normal metal – superconductor ($N-S$) boundary in close proximity to the center of a magnetic curl. Therefore, there is a difference of the electrochemical potentials between the normal metal and the superconductor for the electrons (holes) on the Fermi surface in $YBa_2Cu_3O_{7-\delta}$. This difference in the electrochemical potentials is equal to the condensation energy of the superconducting state, i.e. to the value of the superconducting energy gap $\Delta$. The diffusion transition of an oxygen anion $O^{2-}$ toward the $N \rightarrow S$ direction through the $N$-$S$ boundary (accompanied by the transition of oxygen vacancy in the opposite direction) reduces in the appropriate charge trans-port together with a particle. At this transition, the two holes in the conductivity zone move out the Fermi surface of a normal metal and reach the Fermi surface of a superconductor, where they transit into the superconducting condensate with the particular temperature-dependent probability. This process of diffusion transport of charged particles is accompanied by the addition of electron energy, which is equal to $\sim 2\Delta(T)$. We propose that the presence of the gradient of an electrochemical potential should result into an appearance of the diffusion similar to the so-called chemical diffusion, which is observed in conditions, when there is a gradient of a chemical potential. This process should be accompanied by a decrease of the oxygen index of the normal core of an Abricosov magnetic vortex, and lead to an appearance of an effective centre of pinning, located at Abricosov magnetic vortex core center, which prolates along all the magnetic curl. This additional pinning force, acting on the magnetic vortical structure, results in an increase of critical current density in $YBa_2Cu_3O_{7-\delta}$ thin film samples, depending on the geometry and layout of pinning centers. As it will be shown in the discussion of experimental results, in a number of published research papers, the different authors also researched the change of physical properties of high-temperature superconductors in a magnetic field at nitrogen temperatures, which is probably connected with the modification of oxygen contents level in the oxygen subsystem in high temperature superconductors (*HTS*), caused by the presence of oxygen diffusion processes at the specified conditions. This research aims to find the unified theoretical mechanism to explain the nature of oxygen diffusion, trying to clarify the associated complexities with the oxygen diffusion physical problems, in high-temperature superconductors.

**Experimental settings and results**

The thin superconducting films of $YBa_2Cu_3O_{7-\delta}$ were deposited on the single crystal substrate of $LaAlO_3$ of (100) orientation, using the laser ablation method from the spinned target. The $YBa_2Cu_3O_{7-\delta}$ thin film width was $400\ nm$. After the deposition, the $YBa_2Cu_3O_{7-\delta}$ thin films were processed with the use of annealing process in the oxygen atmosphere within several hours at temperature $T \approx 700K$ with the purpose to increase an oxygen index. The characteristic temperature of superconducting transition in $YBa_2Cu_3O_{7-\delta}$ thin films, measured on the resistance and on the magnetic susceptibility, was equal to $89K$.

The nonequilibrium distribution of oxygen atoms and vacancies was formed by the exposure of superconducting thin films to the electrons irradiation with the energy of $2\cdot10^3\ eV$ up to the maximal fluence $N=5\cdot10^{21}\ electrons/m^2$ at temperature of $77K$. The exposure of $YBa_2Cu_3O_{7-\delta}$ thin films to electrons irradiation was performed at the different currents and fluencies with the durations from several tens of minutes till several hours. In the process of *HTS* thin film exposure to the electron radiation, the $YBa_2Cu_3O_{7-\delta}$ thin films were placed in the magnetic field $B \sim 3\cdot10^2\ G$. The application of magnetic field resulted in an appearance of Abricosov magnetic vortices in $YBa_2Cu_3O_{7-\delta}$ thin film sample. In our case, these Abricosov magnetic vortices served as the centers of gradient of an electrochemical potential, which had an influence on the diffusion transport of the oxygen ions and their vacancies in $YBa_2Cu_3O_{7-\delta}$ thin film. After the exposure process, the not warming capsule with $YBa_2Cu_3O_{7-\delta}$ thin film sample moved into the inductance bridge with the synchronous detection at low frequencies of tens of *Hz*, where the measurement of dependence of the complex magnetic susceptibility $\chi=\chi'+i\chi''$ on the temperature $\chi(T)$ was conducted.

The $\chi'$ real and $\chi''$ imaginary parts of magnetic susceptibility were selected, using the method of the synchronous phase-sensitive detection, shown in Fig. 1. The magnitude of variable magnetic field $H$, acting on the thin film sample, did not exceed $400\ A/m$, and was changed in predetermined manner. The characteristic limiting signal level, accessible for the analysis in the experimental setup, was of several nanovolts.

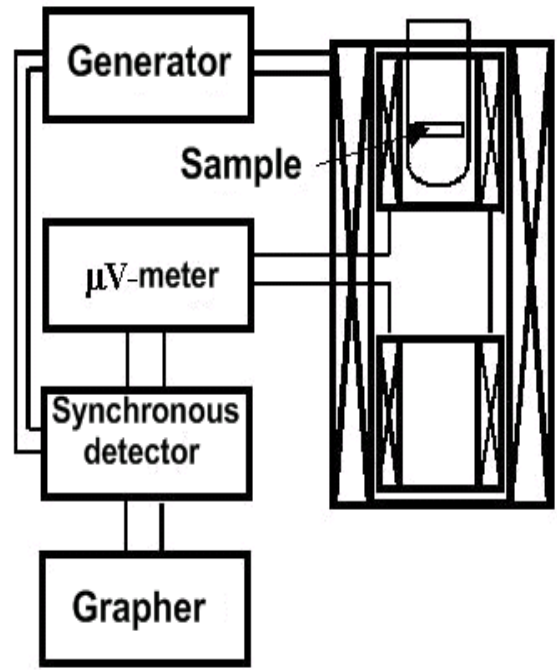

*Fig. 1. The flowchart of experimental setup for the registration of complex magnetic susceptibility and its temperature dependence $\chi(T)$.*

The effective pinning centers of Abricosov magnetic vortices in superconductors are represented by the linear derivations such as the kernels of dislocations, or, for example, by the defects, formed after the passing of heavy particles with high energy in crystal (columnar-defects). In the researched case, the similar effective linear defects appeared during the operation of the oxygen diffusion near to the centers of Abricosov magnetic vortices normal cores during the exposure of $YBa_2Cu_3O_{7-\delta}$ thin film samples to the electron irradiation.

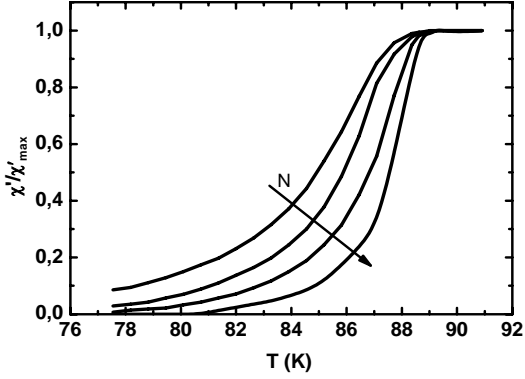

*Fig. 2. Change of real part of relative magnetic susceptibility $\chi'/\chi'_{max}$ in dependence on temperature at increase of fluence N in the irradiated $YBa_2Cu_3O_{7-\delta}$ thin films (the curves located from left to right in accordance with increase of fluence N, the signal frequency is 79 Hz).*

The temperature dependence of a real part of the magnetic susceptibility $\chi'(T)$, obtained during the research on $YBa_2Cu_3O_{7-\delta}$ thin films, is shown in Fig. 2. As it is visible from the graphics, the curve $\chi'(T)$ is shifted in the range of higher temperatures $T$ in accordance with the rise of the fluence $N$ from $1 \cdot 10^{20}$ electron/m² up to $5 \cdot 10^{21}$ electron/m².

Thus, the critical temperature of $YBa_2Cu_3O_{7-\delta}$ thin film sample varies a little, whereas there is a change of a sort of curves that corresponds to the influence of electron irradiation on the modification of order in the oxygen subsystem in the localized areas of crystal grating.

The dependence of the imaginary part of magnetic susceptibility on the temperature $\chi''(T)$ is directly bound to the energy losses, attributed to the penetration of the variable magnetic field and currents into the $YBa_2Cu_3O_{7-\delta}$ thin film sample, resulting in a change of dependence of the maximum of $\chi''(T)$ on the temperature in accordance with increase of fluence $N$ under the exposure of $YBa_2Cu_3O_{7-\delta}$ thin films to the electron irradiation as it is shown in Fig. 3.

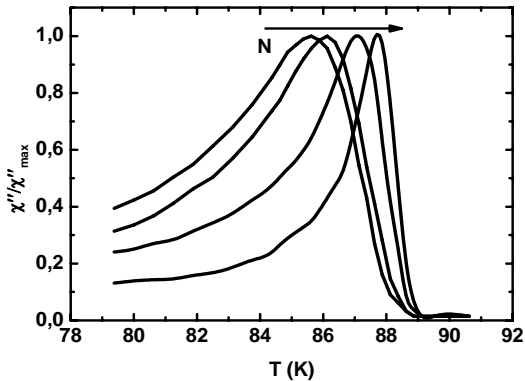

*Fig. 3. Shift of a maxima of imaginary part of relative magnetic susceptibility $\chi''/\chi''_{max}$ in dependence on temperature in accordance with increase of fluence N under the exposure of $YBa_2Cu_3O_{7-\delta}$ thin films to the electron irradiation (the curves are located from left to right in accordance with increase of fluence N, the signal frequency is 79 Hz).*

As it is visible from Fig. 3, the exposure of a $YBa_2Cu_3O_{7-\delta}$ thin film sample to the electron irradiation results in an offset of maxima of dependence of the imaginary part of magnetic susceptibility on the temperature $\chi''(T)$ in the range of higher temperatures $T$. The influence of fluence $N$ on this offset of maxima of dependence of the imaginary part of magnetic susceptibility on the temperature $\chi''(T)$ is shown in Fig. 4. In other words, the dependence of relative change of temperature peak, in the characteristic dependence of the imaginary part of magnetic susceptibility on the temperature $\chi''(T)$, as a function of logarithm of fluence $N$ of electrons, normalized to $10^{20}$ electron/m² at of a $YBa_2Cu_3O_{7-\delta}$ thin film sample to the electron irradiation is constructed in Fig. 4. In the field of small fluencies $N$, the exposure to the electron irradiation has not enough influence on the curve $\chi''(T)$. The curve $\chi''(T)$ quits in the saturation in the field of big fluencies $N$.

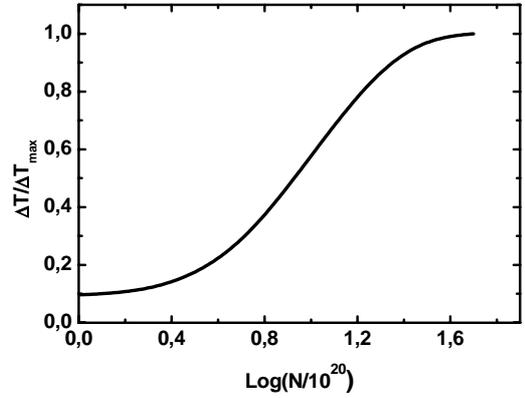

*Fig. 4. Dependence of relative shift of temperature peak $\chi''(T)$ on logarithm of fluence N of electrons, normalized to $10^{20}$ electron/m².*

The saturation of curve of the dependence of an imaginary part of magnetic susceptibility on the temperature $\chi''(T)$ at the big values of fluence $N$ is probably interlinked with the re-allocation of oxygen, which occurs in the nanoscale domains inside the crystal grating in a $YBa_2Cu_3O_{7-\delta}$ thin film sample, making an essential contribution to the increase of Abricosov magnetic vortices pinning, but having a little influence on such macroscopic parameter as the critical temperature of a superconductor $T_C$. It also leads to the situation, when the maxima of the dependence of the imaginary part of magnetic susceptibility on the temperature $\chi''(T)$ reaches a natural limit at the further increase of fluence $N$ at approach to the critical temperature of superconductor $T_C$.

As it follows from the obtained experimental results, the exposure of the $YBa_2Cu_3O_{7-\delta}$ thin films to the electron radiation of rather small energy in a magnetic field $B$ results in the diffusion modification of the oxygen order and composition near to the normal core of Abricosov magnetic vortex. This process has an effect on the shape of dependence of the magnetic susceptibility on the temperature $\chi(T)$ in $YBa_2Cu_3O_{7-\delta}$ thin film.

**Discussion of results**

The diffusion transformation in the oxygen subsystem of a high-temperature superconductor at the temperature of *77 K* can only be realized in the case, when the characteristic activation energy of diffusion process is equal to $E_A \sim 0.1\ eV$. The activation energies of such order of magnitude are characteristic for the diffusion processes observed in the normal metals after their exposure to the electron radiation in the indicated range of temperatures [6].

Let's comment on a number of features of the oxygen diffusion in the high-temperature superconductors. The dependence of the coefficient of diffusion on the relative oxygen index $\delta$ in $YBa_2Cu_3O_{7-\delta}$ high-temperature superconductors is well described by the function $\log(D_{7-\delta}/D_{\delta=0}) \approx -5\delta$ at *720 K* [2]. Therefore, the oxygen deficient crystal grating with the relative oxygen index $\delta = 0.6$ has in $10^3$ times smaller coefficient of diffusion $D_{6.4} \approx 5 \cdot 10^{-13}\ m^2/sec$ than the crystal grating with the relative oxygen index $\delta \sim 0$. As we can see, it confirms the marked above dependence of the magnitude of diffusion coefficient on the contents of oxygen in $YBa_2Cu_3O_{7-\delta}$ crystal grating. The activation energy, which is characteristic for the diffusion processes in $YBa_2Cu_3O_{7-\delta}$ compound, as it is described in different research papers, is $E_A \sim 0.9 \div 1\ eV$ in the **ab** plane. The diffusion coefficient *D* has a sharp non-isotropic dependence on the crystallographic directions, because of the physical property of material.

The smallest value of diffusion coefficient *D* corresponds to the transport of oxygen along the **c** axis with the activation energy $E_A \approx 2.8\ eV$, whereas the biggest value of diffusion coefficient *D* is registered at diffusion process along the **b** axis. There is a little bit smaller diffusion coefficient *D* along the **a** axis than along the **b** axis. Thus, the maximal value of anisotropy of the ratios of the diffusion coefficients reaches the magnitudes $D_a/D_c \approx 10^4$ and $D_b/D_c \approx 10^6$ correspondingly. It is probably connected with the oxygen properties to fill the crystallographic positions in $YBa_2Cu_3O_{7-\delta}$ crystal grating. Thus, in the crystal with the oxygen index close to *7*, all the *O(5)* positions along the **b** axis are unengaged, but the *O(1)* positions are almost completely filled, forming a chain of atoms of oxygen. There should be very small activation energy $E_A$, needed to transport the oxygen and its vacancies toward a direction of the **b** axis, which could be essentially smaller than the activation energies $E_A$ along other axes [3]. In accordance with the increase of the relative oxygen index $\delta$, the oxygen ions at the *O(1)* positions start to take places at the *O(5)* positions partially, and there is a decrease of a value of diffusion coefficient *D*, which is observed in this material. Such structure of $YBa_2Cu_3O_{7-\delta}$ crystal grating facilitates the oxygen diffusion in the **ab** plane, because of the "vacancy-ion" chains presence at the *O(5)* positions, which is already created in the crystal grating, hence the expenditures of energy on their creation is not required [7, 8]. Probably, the anisotropy of oxygen diffusion determines the preferred growth direction of crystals in the **ab** plane. During the synthesis process, the $YBa_2Cu_3O_{7-\delta}$ crystal grows along the **c** axis by microns, whereas its size reaches several millimeters in the indicated **ab** plane. This fact confirms the above mentioned anisotropy of diffusion indirectly.

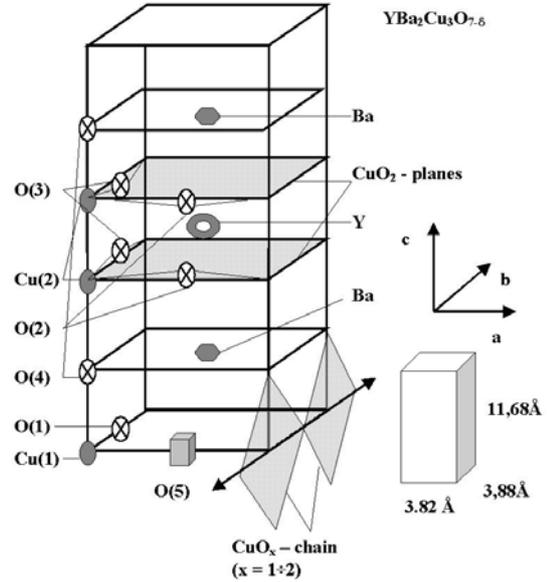

*Fig. 5. Crystalline structure and layout of atoms in a unit cell of $YBa_2Cu_3O_{7-\delta}$ superconductor.*

The diffusion coefficients of other chemical elements, which create the composition of $YBa_2Cu_3O_{7-\delta}$ crystal, have much less values than in the case of the oxygen and its vacancies. Let's note that, even at room temperatures, the noticeable change of oxygen contents is observed in $YBa_2Cu_3O_{7-\delta}$ crystal over period of a few tens of hours. At the same time, at lower temperatures *T*, even at not very big activation energy $E_A \approx 0.4\ eV$, it is not possible to calculate, by virtue of an exponential dependence of the diffusion coefficient on the temperature $D = D_0 exp(-E/kT)$, that the diffusion coefficient of oxygen may have observable value at the temperatures below the critical temperature $T_C$. Let's note that the activation energy of diffusion $E_A$ includes, as a rule, the two contributions:

i. The energy of appropriate defect creation (interstitial atom, vacancy, etc.), and

ii. The activation energy of defect transportation. These processes can be separated, for example, the defects can be generated by the irradiating beam of particles in the case of external exposure of a sample to the irradiation, but the transport processes remain thermoactivated. In this case, the value of activation energy of diffusion can appear rather small: $\sim 0.1\ eV$. For this reason, the diffusion of defects begins already at heating temperatures up to $T \sim 20\text{-}30\ K$ after the exposure to the electron irradiation at the low-temperature of *4.2 K* [6]. The exposure of samples to the electron irradiation with sufficient energy to relocate the atoms, when the excessive vacancies and interstitial atoms can be generated, is used for the creation of indicated dot defects in crystal grating. According to a well known dependence, the limiting transmitted energy for the nonrelativistic particles is equal to

$$E_p = \frac{4EMm}{(M+m)^2}, \qquad (1)$$

where $E$ is the energy and $m$ is the mass of an incident particle, $M$ is the mass of atom in crystal grating [6].

Let's note that the possible ionization mechanisms of excitation appear to be not effective in the normal metals, usually, because the conduction electrons are fast enough to neutralize the nonequilibrium charges for the time $\sim 10^{-16}$ sec. In $YBa_2Cu_3O_{7-\delta}$ materials, this statement is not so clear. At first, the density of electrons is small in comparison with typical normal metals. Secondly, the electronic structure of crystal grating is such that, probably, the planes $CuO_2$ and the chains along the $O(1)$ positions are metallic only, whereas the conductivity along the **c** axis is tunneling. Therefore, the ionization defects, located outside of indicated planes, can have the long time of relaxation. In the oxygen-poor compounds, it appears especially brightly that the conductance of the indicated planes decreases strongly. In this case, the exposure of crystals to the light of sufficient intensity reduces in the subsequent long-lived relaxation of their electron conduction with the activation energy $E_A \sim 0.93$ eV [9], which is close to the activation energy $E_A$ of a diffusion of oxygen in the **ab** plane. The presence of oxygen anions, which are ionized by the electron beam in different positions in crystal grating, can have an effect on their more intensive transport, stipulated by the fields of electrochemical potentials of defects at vacancy positions, which already are available in crystal grating of $YBa_2Cu_3O_{7-\delta}$. It is clear that the influence of low energy electrons with $E \sim 2 \cdot 10^3$ eV, researched in the present work, can only be observed in the thin film systems in view of both the small penetration depth of electron irradiation and the small value of energy transmitted to the atoms by the electrons. Thus, the higher energy, according to eq. (1), is gained by the atoms with small masses, i.e. the atoms of oxygen in $YBa_2Cu_3O_{7-\delta}$ superconductor. The characteristic limiting energy, transmitted in the oxygen subsystem, appears to be equal to $\sim 3$ eV per a transportable atom of oxygen. This value of energy is enough to create a nonequilibrium population of free $O(5)$ positions by the oxygen atoms, coming from other positions, this mechanism reduces in the origin of a diffusion process with the small activation energy $E_A \sim 0.1$ eV in $YBa_2Cu_3O_{7-\delta}$ crystal. In result, probably, the intensive diffusion transport of oxygen and its vacancies can already have place in the $YBa_2Cu_3O_{7-\delta}$ compounds at temperature of 77 K.

Let's consider the contribution by the superconductivity to the diffusion process in details. It is known that the transition to the superconducting state can have an impact on the mechanical properties of crystal grating of superconductors. It is necessary to state that the characteristic energy, which is connected with the superconducting transition and corresponds to the diffusion transport of one ion in crystal grating, has the magnitude $10^{-7}$ eV in low-temperature superconductors. Thus, the characteristic energy is distributed among all the crystallographic positions in space homogeneously, and reduces in a change of volume of a single cell of crystal grating on the rather small value about $10^{-7}$ of relative volume. As it was mentioned above, the defects of structure can be created at an expense of energy of an incident beam during the exposure of crystal grating to the electron irradiation, and the diffusion coefficient of defects is only determined by the transport energy and can be rather small. Thus, under the exposure of the copper $Cu$ to the electron irradiation at low-temperature, the activation energies $E_A$ of diffusion processes start at hundredth of the eV at low temperatures and reach about 0,1 eV at the temperatures of 70 K [6]. As we can see, the mobility of defects is sufficient for the change of allocation of defects in $Cu$ even at low temperatures. When the dislocations are available in a crystal grating, such mobility reduces in the capture of defects by the deformation field of a dislocation, and there is a cloud formed by the defects and atoms of impurities (Kotrell's cloud), which is enclosing it. This allocation of the defects and impurity atoms, as it is known, anchors the dislocations in the crystal grating, restricting their mobility, and increments the ultimate strength of material [10]. Thus, the characteristic gradient of electrochemical potential, acting on the impurity, reaches the magnitude of $10^7$ eV/m.

This fact is rather interesting, when considering the Abricosov quantized vortical magnetic lines in a high-temperature superconductor, because it would lead to the increase of density of critical current in a superconductor, if the Abricosov quantum magnetic curls could be anchored in the same way as in the case of the dislocations, and their pinning could be increased. However, the rotational magnetic curls have no sufficient magnitude of deformation field, and consequently can not effectively interact with the impurities and defects by the means of the deformation mechanism.

The condensation energy of electrons reduces the thermodynamic potential of crystal grating, if it is in a superconducting state. Generally, as it was mentioned above, it appears in both the change of size of a superconductor in comparison with the normal state, and the change of its elastic properties. In close proximity to the Abricosov magnetic vortex, we have to consider the locality of the normal core of an Abricosov magnetic vortical quantum line. As it is well known, the Abricosov magnetic vortex core is in the state, which is close to the normal metal state with the strongly suppressed superconducting gap $\Delta$, spreading on the distance of the superconducting correlation length $\xi$ from the center of the normal core of Abricosov magnetic vortex approximately. Further, the circulating superconducting currents flow around the vortical line on the distances up to the $\lambda$ penetration depth of the magnetic field $B$. Thus, the magnetic quantum line has certain symmetry, and there are no essential singularities of deformation fields, as it is observed in the case of a dislocation. However, there is a difference of electron energies, which leads, as we will demonstrate below, to the same effect approximately.

According to the research findings, the superconductivity of $YBa_2Cu_3O_{7-\delta}$ crystal grating is mainly determined by the superconductivity of the $CuO_2$ planes. The oxygen subsystem of base plane of crystal grating, where there are oxygen vacancies in the $O(5)$ positions, even, when the oxygen index of crystal grating is equal to *7*, essentially, has a role of supplier of the hole states, which influence the superconducting properties of $CuO_2$ planes. In $YBa_2Cu_3O_{7-\delta}$ crystal the doping role is played by the atoms and vacancies of oxygen, which are located in the *O(1)* и *O(4)* positions in Fig. 5. The outgo of oxygen atoms from the *O(1)* position reduces the density of electronic states in the $CuO_2$ plane, and correlates with the decrease of critical temperature $T_C$. Therefore, the migration of oxygen and its vacancies leads to the local change of density of electron states. The condensation energy, corresponding to the one superconducting pair of electrons, amounts from *20 meV* up to *50 meV*. Though, in a general case, the electronic states in a crystal grating are grouped, and have the uniform Fermi surface, we can allocate the Abricosov magnetic vortex core, which is in a normal state, and consider the transition of the oxygen atom or vacancy through the *N-S* boundary. This transition is similar to the well known transitions of an electron state through a barrier, dividing the normal and superconducting phases, i.e. the *S-N-S* Josephson junction, and affects the states located close to the Fermi surface only. It is accompanied by the annihilation or creation of superconducting electron pair. The contribution of such transition to the thermodynamics of a crystal grating is equal to the condensation energy of superconducting electron pair. We consider not only the electron transitions through a barrier, but the transition of the oxygen atom or vacancy together with the electron excitations through the normal metal – superconductor boundary without the barrier, and consequently, the diffusion of oxygen atom is quite possible, and it is not suppressed by a barrier. At transition through the *N-S* boundary, the electron excitations can move in or out on the Fermi energy level only, and in the case of a superconductor, they take part in the creation of superconducting condensate; but in the case of normal metal, they participate in the electron transport simply. Therefore, the transition of oxygen atoms or its vacancies through the *N-S* boundary is accompanied by the change of a number of electron pairs in a superconducting state, since there is, as in the Josephson effect, the annihilation or creation of electronic states on the Fermi surface in a superconductor. The thermodynamic potential varies on the value of the energy of superconducting gap *Δ*. Now, let us provide a quantitative assessment of the considered phenomena and its characteristic transition time.

Let's write the diffusion equation, corresponding to the considered research problem, and pay attention to the fact that in this case, it is necessary to take into the account the existence of a gradient of an electrochemical potential. Let's suppose that the transport of an atom is connected with the transport of one electron state, though the oxygen is divalent generally. In our case, this circumstance has no key value and can only amplify the considered phenomena. The diffusion equation can be written as the general expression

$$\frac{\partial C}{\partial t} = -D\frac{\partial}{\partial r}\left(\frac{\partial C}{\partial r} + \left(\frac{C}{kT}\right)\frac{\partial \Delta}{\partial r}\right) \quad (2)$$

where *C* is the concentration, *D* is the diffusion coefficient, *Δ* is the superconducting energy gap, *k* is the Boltzmann constant, *T* is the temperature, *r* is the distance, *t* is the time. We are interested in that part of the equation, which is connected with the dependence of the superconducting energy gap *Δ* on the distance *r*, therefore, we will obtain

$$\frac{\partial C}{\partial t} = -\left\{\frac{D}{kT}\left(\frac{\partial C}{\partial r}\right)\frac{\partial \Delta}{\partial t} + D\left(\frac{C}{kT}\right)\frac{\partial^2 \Delta}{\partial r^2}\right\} \quad (3)$$

Let's take into the account the fact that the concentration of oxygen in the initial moment does not depend on the distance *r*, and then this equation will look like

$$\frac{\partial C}{\partial t} = -D\left(\frac{C}{kT}\right)\frac{\partial^2 \Delta}{\partial r^2} \quad (4)$$

At the following stages of process, when the oxygen composition in Abricosov magnetic vortex core is varying, it also includes the terms, which are connected with $\partial C/\partial r$.

The stream of atoms of oxygen and its vacancies can be determined from the Fick's equation, which may be written with the consideration of dependence of the superconducting energy gap *Δ* on the distance *r*

$$J_O = A \cdot D \frac{\partial C}{\partial r} = A \cdot D \frac{\partial C}{\partial \Delta}\frac{\partial \Delta}{\partial r}, \quad (5)$$

where *A* is a constant.

The local superconducting energy gap *Δ* depends on the concentration of oxygen in $YBa_2Cu_3O_{7-\delta}$ crystal grating. Let us accept the value of the oxygen index *7* as a unit (*C=1*) for the concentration of oxygen, and taking into the account the fact that the superconducting energy gap *Δ* decreases from $2 \cdot 10^{-2}$ *eV* to zero at the decrease of the oxygen index to *6,5* (*C≈0,93*), we will get $\partial C/\partial \Delta = 3,57 \ eV^{-1}$ in the linear approximation. Taking into the account the fact that the main change happens on the correlation length $\xi \approx 2 \cdot 10^{-9}$ *m*, the value of a gradient of superconducting energy gap *Δ* will be equal to $\partial \Delta/\partial r \approx 10^7 \ eV/m$. The leap of thermodynamic potential has different signs in the case of the oxygen atoms and the vacancies, and therefore, they move through the *N-S* boundary in the opposite directions. The diffusion coefficient is

$$D = D_0 exp(-E_A/kT), \quad (6)$$

where $D_0 \approx 2,6 \cdot 10^{-8} \ m^2/sec$, and we will have $D \approx 1,625 \cdot 10^{-14} \ m^2/sec$ at the characteristic activation energy $E_A \approx 0,1$ *eV* and the temperature of *77 K*. Then, the characteristic time of oxygen diffusion on the correlation length ξ in close proximity to the Abricosov magnetic vortex normal core appears to be $\tau = \xi^2/D \approx 2,46 \cdot 10^{-4}$ *sec*. We will obtain the diffusion coefficient $D \approx 5,83 \cdot 10^{-22} \ m^2/sec$ and the characteristic time of oxygen diffusion $\tau = 400 \ sec$ for the activation energy $E_A = 0,2 \ eV$.

Going from these estimations, it is evident, that the transport of the oxygen ions and its vacancies at the operation of a gradient of superconducting potential happens quite fast, and the process is similar to the well known chemical diffusion. The only difference, here, is in the fact that the leap of electrochemical potential on the *N-S* boundary is stipulated by the superconductivity. Probably, this theoretical mechanism is also responsible for some other physical properties, which were earlier observed in the superconductors in the magnetic field in some researches by other authors. In [11], during the research on the internal friction, it was revealed that there was an irreversible modification of a crystal grating in bismuthic ceramics at nitrogen temperature at presence of the magnetic field $H \sim 1.6 \cdot 10^4 A/m$. The annealing of a sample at high enough temperatures was required for the subsequent elimination of the irreversibility, and consequently, it is possible to assume that the irreversible modification has the diffusion nature, connected with the diffusion transport of oxygen and its vacancies near to the Abricosov magnetic quantum lines, considered in the given research. After the heating of a sample up to the room temperature, the derived linear physical properties of the composition of crystal grating continued to appear in the range of high enough temperatures. The change of melting line of a lattice of Abricosov magnetic vortices was observed [12] after the exposition of a crystal grating to the magnetic field at different periods of time. Probably, such expositional influence, is also connected with the processes of pinning of Abricosov magnetic vortices by the oxygen diffusion in a superconductor at temperature $T < T_C$. The decay and appearance of normal non-superconducting domains in a crystal of high-temperature superconductor in a superconducting state under the operation of high density current were reported in [13]. The direct experiment on the $YBa_2Cu_3O_{7-\delta}$ film samples was conducted in [14], in which the current pulses, accompanied by the generation of voltage $\sim 10^5$ $V/m$ in the place of contraction, resulted in the diffusion transport of the oxygen and its vacancies on a few microns distance during a few seconds time period. This process was observed by a method of the micro-Raman laser spectroscopy, and was considered as a confirmation of possibility of diffusion transport of the oxygen atoms and its vacancies under the operation of a gradient of an electrochemical potential.

**Conclusion**

In this work, the research on the accurate characterisation of superconducting properties of $YBa_2Cu_3O_{7-\delta}$ thin films, exposed to the low energy electron irradiation with the energy of $2 \cdot 10^4$ $eV$ are conducted. The experimental measurements to accurately characterize the $YBa_2Cu_3O_{7-\delta}$ thin films in the magnetic field of $H \sim 2.4 \cdot 10^4$ $A/m$ at the temperature $T = 77$ $K$ are completed. The phenomena of oxygen and its vacancies diffusion near to the normal cores of Abricosov magnetic vortices in $YBa_2Cu_3O_{7-\delta}$ thin films is discovered and researched. It is shown, for the first time, that the diffusion of the oxygen ions and its vacancies near to the Abricosov magnetic quantum lines is connected with the gradient of an electrochemical potential, in which the superconducting contribution is significant and must be taken to the consideration. It is demonstrated that the oxygen deficient domain is created in the Abricosov magnetic vortex normal core, representing an effective pinning center. This additional pinning centers results in an increase of the density of critical current in $YBa_2Cu_3O_{7-\delta}$ thin films. The introduced theoretical mechanism consistently explains the phenomena of the oxygen and its vacancies diffusion in $YBa_2Cu_3O_{7-\delta}$ thin films near to the Abricosov magnetic quantum lines, adding some clarity to a number of early obtained experimental results, which had no clear explanation before.

The authors express thanks to I.N. Chukanova for the synthesis of $YBa_2Cu_3O_{7-\delta}$ thin films. Viktor O. Ledenyov thanks K. Alex Müller for the thoughtful discussions on the oxygen diffusion in $YBa_2Cu_3O_{7-\delta}$ at *NATO ASI* in Loen, Norway in 1997; and appreciates K.C. Goretta for the presentation materials and detailed discussion on the atomic transport in *HTS* at *Leonardo da Vinci IAS* in Bologna, Italy in 1998.

This research paper was published in Russian in the Problems of Atomic Science and Technology (*VANT*) [15].